\documentclass{pasj00}
\Received{2014 September 1}
\Accepted{2014 October 21}
\Published{$\langle$in press$\rangle$}
\SetRunningHead{Obscured NLS1 Candidate Mrk~1388}{Akihiro Doi}

\usepackage{times}
\usepackage{natbib}
\newcommand{\nar}{New Astronomy Reviews}

\begin{document}

\title{Obscured Narrow-Line Seyfert 1 Galaxy Candidate Mrk~1388\\ with Nonthermal Jets}
\author{Akihiro \textsc{Doi}}%
\affil{The Institute of Space and Astronautical Science, Japan Aerospace Exploration Agency,\\ 3-1-1 Yoshinodai, Chuou-ku, Sagamihara, Kanagawa 229-8510}
% \affil{Department of Space and Astronautical Science, The Graduate University for Advanced Studies,\\ 3-1-1 Yoshinodai, Chuou-ku, Sagamihara, Kanagawa 229-8510}
% \email{akihiro.doi@vsop.isas.jaxa.jp}

\maketitle

\begin{abstract}
Mrk~1388 has an unusual Seyfert nucleus that shows narrow emission-line components without broad ones, but shows a strong featureless continuum and strong iron-forbidden high-ionization emission lines.  The apparent coexistence of type-1/2 characteristics is potentially attributed to a heavily obscured broad-line region or to an intermediate-mass black hole with a broad-line component intrinsically narrower than those of typical narrow-line Seyfert~1~(NLS1) galaxies.  Our observation using very-long-baseline interferometry~(VLBI) reveals high-brightness radio emission from nonthermal jets from an active galactic nucleus~(AGN) with a significant radio luminosity.  Furthermore, we investigate the radial profile of the host galaxy using a Hubble Space Telescope~(HST) image, which shows a S$\mathrm{\acute{e}}$rsic index suggestive of a pseudobulge.  Using the VLBI and HST results, which are essentially not affected by dust extinction, three individual methods provide similar estimates for the black hole mass: $(0.76$--$5.4)\times10^6 \Mo$, $1.5\times10^6 \Mo$, and $4.1\times10^6 \Mo$.  These masses are in a range that is preferential for typical NLS1 galaxies rather than for intermediate-mass black holes.  Based on the estimated masses, the full width at half maximum $FWHM(\mathrm{H\emissiontype{$\beta$}})$ of approximately 1200--1700~km~s$^{-1}$ should have been seen.  The scenario of a heavily absorbed NLS1 nucleus can explain the peculiarities previously observed.   
\end{abstract}

% ***********************************************************************************
\section{Introduction}\label{section:introduction}
Mrk~1388 has an unusual Seyfert nucleus that shows the characteristics of both type-1 and type-2 Seyfert galaxies (hereinafter Sy1 and Sy2, respectively).  
Permitted lines such as the hydrogen Balmer emission lines of H\emissiontype{$\alpha$} and H\emissiontype{$\beta$} are narrow ($280$ and $340$~km~s$^{-1}$, respectively) with essentially the same widths as forbidden lines at the lower end of the line-width distribution for Seyfert galaxies \citep{Osterbrock:1985a}.  
The ratio of an oxygen-forbidden line and the Balmer line, [O\emissiontype{III}]$\lambda5007$/H\emissiontype{$\beta$}~$=11.1$, is much more typical of Sy2 ([O\emissiontype{III}]$\lambda5007$/H\emissiontype{$\beta$}~$>3$ for a Sy2s criterion).  
The iron emission line of Fe\emissiontype{II} is very weak, as is typical for Sy2s \citep{Osterbrock:1985}.   
On the other hand, \citet{Osterbrock:1985a} observed a strong featureless continuum (as for Sy1) and strong iron-forbidden high-ionization emission lines such as ([Fe\emissiontype{VII}]$\lambda6087$ + [Fe\emissiontype{X}]$\lambda6375$)/[O\emissiontype{III}]$\lambda5007 = 0.044$, which is higher than typical values for Sy1s but lower than the most extreme examples.  
Mrk~1388 was included by \citet{Osterbrock:1985} as a narrow-line Seyfert~1 (NLS1) galaxy on the basis of the strong high-ionization emission lines of [Fe\emissiontype{VII}]$\lambda\lambda5721$, $6087$, and [Fe\emissiontype{X}]$\lambda6375$, whereas \citet{Osterbrock:1984} noted that Mrk~1388 may be most simply described as an unusually high-ionization Sy2 galaxy\footnote{Mrk~1388 is listed as a type-1.9 Seyfert galaxy in the catalogue of quasars and active nuclei: 13th edition \citep{Veron-Cetty:2010}, which refers to \citet{Osterbrock:1985a}.}.  
The current classification for NLS1 galaxies \citep{Pogge:2000} requires (1)~narrow permitted lines only slightly broader than forbidden lines, (2)~$FWHM$(H\emissiontype{$\beta$}$) < 2000$~km~s$^{-1}$, and (3) [O\emissiontype{III}]$\lambda5007$/H\emissiontype{$\beta$}~$<3$; however, exceptions are allowed if strong [Fe\emissiontype{VII}] and [Fe\emissiontype{X}] are also present, unlike what is seen in Sy2 galaxies.  Thus, Mrk~1388 lacks the broad-line property to be classified as an NLS1 galaxy.  
The narrow component Pa\emissiontype{$\beta$} of the hydrogen Paschen emission line is detected in the $J$-band in the near infrared; however, no broad component appears in Pa\emissiontype{$\beta$} \citep{Goodrich:1994,Veilleux:1997}, which suggests a heavily obscured ($A_\mathrm{V} \gtrsim11$~mag) broad-line region~(BLR) or a broad-line component intrinsically narrower than those of typical NLS1 galaxies because of an intermediate-mass black hole ($10^3$--$10^6~\Mo$) for active galactic nucleus (AGN) activity.  

Radio observations are not affected by dust extinction toward the nucleus through our line of sight.  Based on observations by the Very Large Array (VLA) A-array configuration at 1.49 and 4.86~GHz \citep{Ulvestad:1995}, Mrk~1388 is an unresolved radio source of nonthermal emission with a power-law spectrum with an index $\alpha=-0.80$ ($S_\nu \propto \nu^\alpha$, where $S_\nu$ is the flux density at frequency $\nu$).  The standard infrared/radio ratio $q$ \citep{Condon:1995} for Mrk~1388 is significantly small, which suggests that AGN dominates star-forming activity \citep{Dennefeld:2003}.  No polycyclic aromatic hydrocarbon (PAH) feature, which also provide excellent diagnostics to distinguish starburst and AGN energy sources, is detected for Mrk~1388 \citep{Dennefeld:2003}; this is consistent with the low infrared/radio ratio $q$.  The appreciable jet activity is powered by proportionate mass accretion with limits lesser than the order of the Eddington luminosity, which is proportional to the black hole mass.  Thus, radio jet clues to the mass of the central black hole.

The properties of the bulge component correlate with the mass of the central black hole \citep{Ferrarese:2000,Gebhardt:2000}, although they do not correlate with a pseudobulge or galaxy-disk component \citep{Kormendy:2011}.  
Mrk~1388 was described as ``very compact'' \citep{Zwicky:1963}.  \citet{Markaryan:1979} discovered this galaxy and described it as spherical, with diffuse edges, and apparently with a star-like nucleus.  However, its morphology is actually slightly elongated, and thus is elliptical rather than spherical \citep{Osterbrock:1985}.  The Hubble type is ``E?'' in the NASA/IPAC Extragalactic Database (NED); the $24$-mag/arcsec$^2$ radius is only $6.4$~kilo-parsecs~(kpc).  Mrk~1388 is an isolated galaxy with almost no asymmetry or distortion \citep{Xanthopoulos:1991}, suggestive of long since a last merging event.  Thus, the host of Mrk~1388 may be a small and isolated elliptical galaxy.  This peculiar galaxy may provide important clues for extreme-parameter regions connecting bulges and central black holes in the scenario of coevolution.

The present paper discusses the mass of the central black hole in Mrk~1388 based on very-high-angular-resolution radio and optical studies that used the very-long-baseline interferometry~(VLBI) technique and the Hubble Space Telescope~(HST), respectively.   
In Section~\ref{section:dataanddataanalysis}, we describe our VLBI observation and data reduction and analysis of an HST image.  The results are presented in Section~\ref{section:result}, and their implications are discussed in Section~\ref{section:discussion}.  
Throughout this paper, we use Lambda cold dark matter ($\Lambda$CDM) cosmology with $H_0=70.5$~km~s$^{-1}$~Mpc$^{-1}$, $\Omega_\mathrm{M}=0.27$, and $\Omega_\mathrm{\Lambda}=0.73$.  The redshift is $z=0.021169 \pm 0.001011$~(Sloan Digital Sky Survey Data Release~7); the luminosity distance is $101$~Mpc, and the angular-size distance is $96.8$~Mpc; 1~milliarcsecond~(mas) corresponds to a projected linear scale of $0.469$~pc at the distance to Mrk~1388.

% ***********************************************************************************
\section{Data and Data Analysis}\label{section:dataanddataanalysis}
\subsection{VLBI observation}\label{section:VLBAobservation}
We observed Mrk~1388 on May~9, 2005 in the $L$-band using ten antennas of the Very Long Baseline Array~(VLBA) at the National Radio Astronomy Observatory~(NRAO).  The data were obtained with observation code BD106, in which seven objects of nearby radio-quiet NLS1 galaxies were also observed.  The results of these observations were previously published \citep{Doi:2013a}.      
A left-circular polarization was received at a center frequency of $1.667$~GHz with a total bandwidth of $32$~MHz.  Because Mrk~1388 is a faint radio source, we observed in phase-referencing mode, which allowed us to derive calibration parameters for instrumental and atmospheric effects from observations of the nearby strong compact radio source (``calibrator'') \citep{Beasley:1995} J1455+2131 separated by $\timeform{1.66D}$ from the target.  The period of the antenna nodding cycle was $5$~min, and the total on-source time was approximately $40$~min for the target.  
Standard calibration procedures for VLBA phase referencing were applied during data reduction using the Astronomical Image Processing System ({\tt AIPS}; \citealt{Greisen:2003}) software, developed by the NRAO.  The details of data reduction are the same as for \citet{Doi:2013a}.  
We set the phase-tracking center for Mrk~1388 to the position determined using the VLA A-array at $5$~GHz with an accuracy of approximately $\timeform{0".1}$ \citep{Ulvestad:1995}.  An emission was initially found at a position offset by less than several tens of mas.  After shifting the mapping center, we deconvolved the images using the task {\tt IMAGR} (CLEAN), and we measured the astrometric positions of the emission using the task {\tt JMFIT}.  Subsequently, we self-calibrated in phase with a threshold signal-to-noise ratio of $2.5$ using the task {\tt CALIB}.  We smoothed and interpolated the solutions before applying to the data.  The deconvolution and self-calibration algorithms were interactively applied several times.  

\begin{figure}
\begin{center}
\FigureFile(0.9\linewidth, ){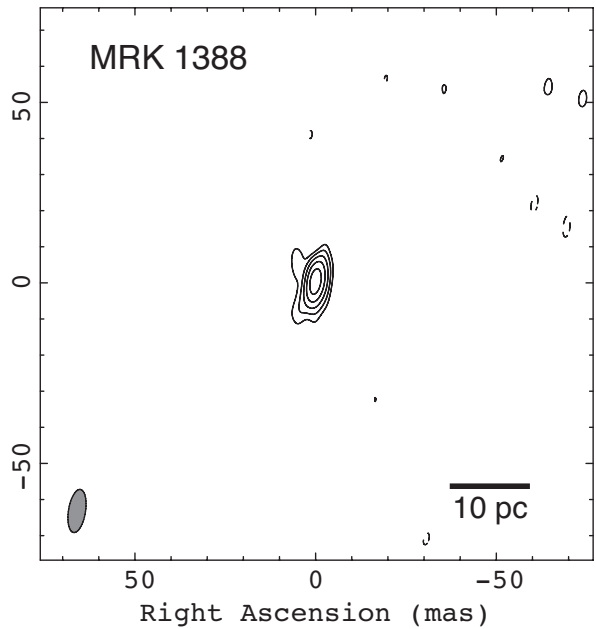}
\end{center}
\caption{VLBA image for Mrk~1388 at $1.667$~GHz in the region of $150 \times 150$~mas$^{2}$.  Contours are separated by factors of $\sqrt{2}$ beginning at three~times the rms noise.  Negative and positive contours are represented by dashed and solid curves, respectively.  The half-power beam size is given in the lower-left corner.  Angular scale corresponding to a linear scale of $10$~pc is given in the lower-right corner.}
\label{figure:VLBA}
\end{figure}

We used the {\tt Difmap} software \citep{Shepherd:1994} to make the final image using the CLEAN algorithm from the calibrated visibilities.  To retrieve both compact and diffuse components to the extent possible, we used step-by-step uniform, natural, and ($u$,~$v$)-tapered weighting functions for CLEAN.  The final VLBA image is shown in Figure~\ref{figure:VLBA}.  Image parameters are listed in Table~\ref{table:VLBA}.

\begin{table*}
\caption{Results of VLBA observation.\label{table:VLBA}}
\begin{center}
\begin{footnotesize}
\begin{tabular}{ccccccccccc}
\hline
\hline
Object&	$I_\mathrm{1.7GHz}$&	$S_\mathrm{1.7GHz}$&	$\sigma$&	$\theta_\mathrm{maj}$&	$\theta_\mathrm{min}$&	$P.A.$&	$\log(T_\mathrm{B}/\mathrm{K})$&	$L_\mathrm{5GHz}$& \multicolumn{ 2}{c}{Astrometric position (J2000.0)} \\
&	(mJy/b)&	(mJy)&	(mJy/b)&	(mas)&	(mas)&	(\timeform{D})&	&	(erg~s$^{-1}$)& R.A.&	Decl.\\
(1) & (2) & (3) & (4) & (5) & (6) & (7) & (8) & (9) & (10) & (11) \\
\hline
Mrk~1388&	$1.8\pm0.2$&	$3.4\pm0.2$&	0.140&	12.1&	4.7&	$-$9.0&	7.6& $8.6\times10^{37}$ & 	\timeform{14h50m37.8497s}&	$+$\timeform{22D44'03.601''}\\
\hline
\end{tabular}
\end{footnotesize}
\end{center}
\begin{flushleft}
\begin{footnotesize}
Col.~(1) Target name; 
Col.~(2) peak intensity; 
Col.~(3) total flux density; 
Col.~(4) image rms~noise on blank sky; 
Cols.~(5)--(7) synthesized beam sizes in major axis, minor axis, and position angle of major axis, respectively; 
Col.~(8) brightness temperature; 
Col.~(9) 5-GHz radio luminosity in rest frame assuming spectral index $\alpha=-0.8$ \citep{Ulvestad:1995}; 
Cols.~(10)--(11) astrometric positions determined by VLBA observation.
\end{footnotesize}
\end{flushleft}
\end{table*}

\subsection{Hubble Space Telescope data}\label{section:HSTdata}
We used an HST archival data set (Dataset = U2E62101P), which contains an the observation obtained August~25, 1994 using a $500$~s exposure with the F606W filter in the field-of-view of Planetary Camera~1~(PC1) in the Wide Field and Planetary Camera~2~(WFPC2).  The archival data were calibrated by the On-The-Fly Reprocessing (OTFR) system developed at the Space Telescope Science Institute.    
The two-dimensional radial profile of Mrk~1388 was analyzed using {\tt GALFIT} \citep{Peng:2002}.  A point spread function~(PSF) was made using the web-based modeling tool Tiny Tim\footnote{http://www.stsci.edu/hst/observatory/focus/TinyTim} with four-times oversampling, which was then downloaded and applied to the GALFIT analysis.  The magnitude Zeropoint was corrected for F606W~($22.887$~mag).  We initially tried modeling with a S$\mathrm{\acute{e}}$rsic $+$ sky bias model, a two-S$\mathrm{\acute{e}}$rsic $+$ sky bias model, and a S$\mathrm{\acute{e}}$rsic $+$ exponential disk $+$ sky bias model, but these combinations failed.  However, a two-S$\mathrm{\acute{e}}$rsic $+$ exponential disk $+$ sky bias model provided a set of plausible solutions (Table~\ref{table:GALFIT}).   

The first and second S$\mathrm{\acute{e}}$rsic components have parameters with S$\mathrm{\acute{e}}$rsic indices $n=1.27$ and $n=0.10$, respectively.    
The latter is significantly resolved ($R_\mathrm{e}=1.45$~pixels) with respect to the PSF.  This component appeared clearly at the center of galaxy in the residual image of the single S$\mathrm{\acute{e}}$rsic case; if we use a PSF component instead we cannot represent this central excess.      
The exponential disk has the scale length $R_\mathrm{s}=1.1$~kpc.  Figure~\ref{figure:HST} shows the radial profiles of the data and the fits.

\section{Result}\label{section:result}

The detected radio emission of Mrk~1388 is comparable or lower than radio powers emanated from the most radio-luminous starbursts, which are approximately $10^{22.3}$--$10^{23.4}$~W~Hz$^{-1}$ at $5$~GHz (equivalent to $10^{39}$--$10^{40.1}$~ergs~s$^{-1}$) \citep{Smith:1998}.    
However, the VLBI detection gives a brightness temperature of $10^{7.6}$~K (Table~\ref{table:VLBA}), which is too high to attribute to any stellar origin.  We can conclude, therefore, that the detected radio emission is relevant to the activity of AGN, which is presumably nonthermal synchrotron jets on the pc or sub-pc scales.      

No clearly elongated structure suggestive of jets is evident in the limited dynamic range of the image (Figure~\ref{figure:VLBA}).  However, the deconvolution shows a significantly resolved structure, which is also supported by the significant discrepancy between peak intensity and integrated flux density (Table~\ref{table:VLBA}) in the self-calibrated VLBA image.   
The VLBA retrieved $43$\% of the expected 1.7~GHz total flux density estimated from a 1.4~GHz VLA flux density of $9.1$~mJy and a spectral index of $\alpha=-0.80$ \citep{Ulvestad:1995}.  The remaining portion indicates the presence of low-brightness components with linear scales greater than $10$~pc, which presumably originated in extended AGN jets and/or in supernova remnants.  
Thus, we confirm that, in the radio regime, approximately half or more of the energy source in Mrk~1388 originates in the AGN.  Previous suggestions of the predominance of the AGN compared with star-forming activity, which were based on the low infrared/radio ratio $q$ and the negative detection of PAH features \citep{Dennefeld:2003}, are confirmed by our VLBI observation.

\begin{figure}
\begin{center}
\FigureFile(0.9\linewidth, ){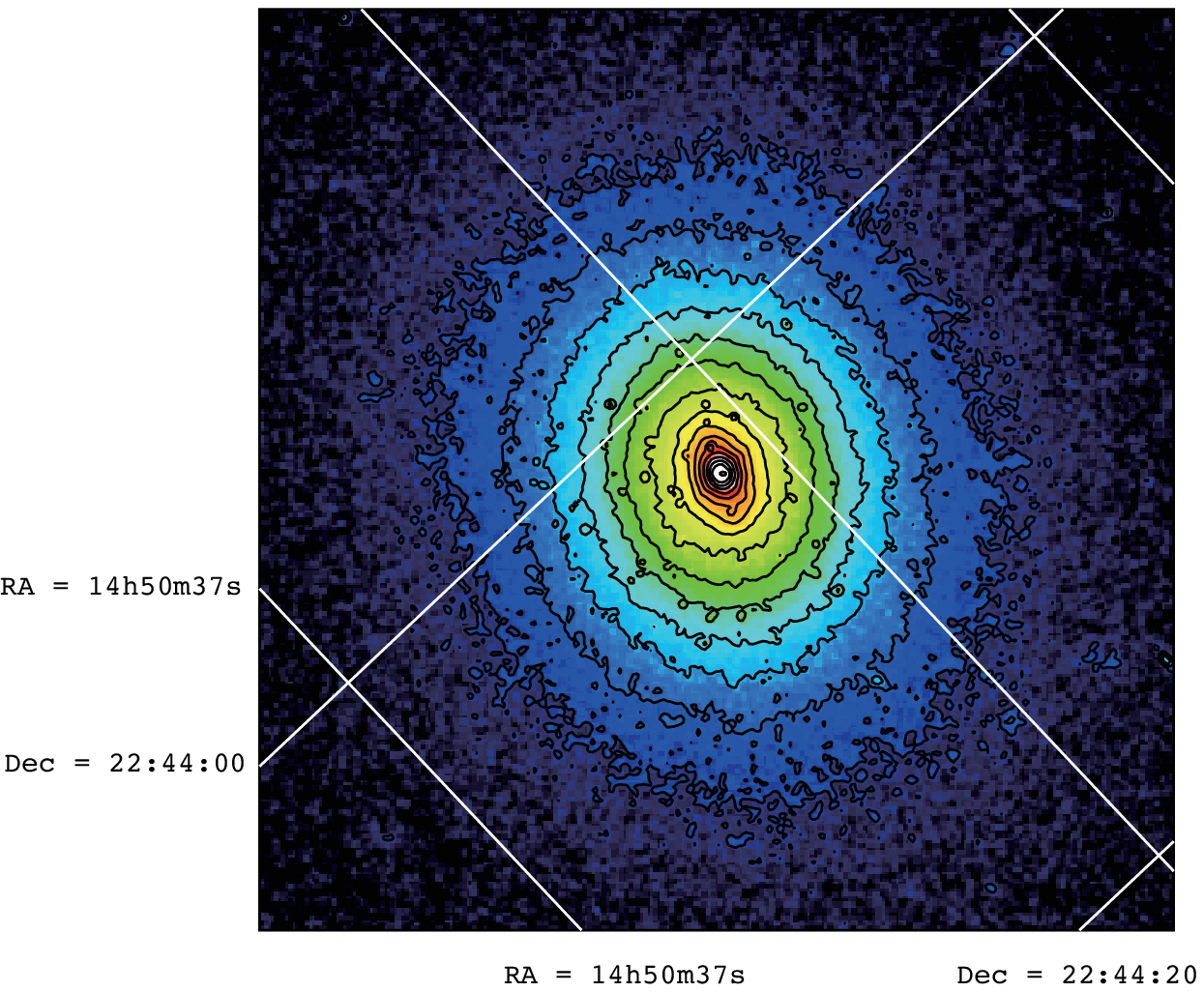}
\FigureFile(0.9\linewidth, ){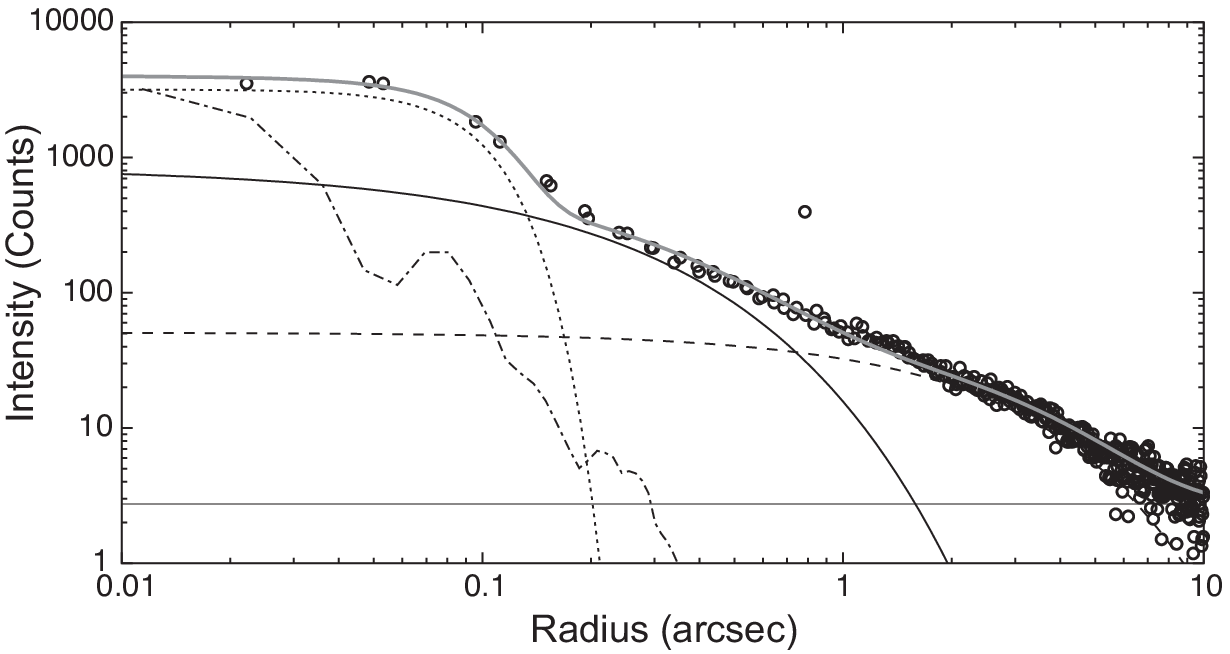}
\end{center}
\caption{Top panel shows HST image of Mrk~1388.  The image is oriented at $-\timeform{136.69459D}$ from North.  Bottom panel shows slice profile along semi-major axis of bulge component.  Open circles represent pixel data in counts in a pixel size of \timeform{0.0455''}.   
Solid curve represents bulge component of a function fit with a S$\mathrm{\acute{e}}$rsic index of $n=1.27$ and an effective radius of $R_\mathrm{e}= 0.23$~kpc.  
Dotted curve represents nuclear star cluster of a fitted function with $n=0.10$ and $R_\mathrm{e}= 0.031$~kpc.  Dashed curve represents a fit to an exponential disk with a scale length of $R_\mathrm{s}=1.1$~kpc.  Thin solid line represents a sky bias fit.   
Gray solid curve represents the sum of them of dotted, dashed, solid, and thin solid curves.  As a reference, the dot-dashed curve represents PSF used for deconvolution (no PSF component was used in GALFIT).}
\label{figure:HST}
\end{figure}

The GALFIT modeling determined well the two-dimensional profile of the host galaxy of Mrk~1388 (Figure~\ref{figure:HST}).  The notable feature characterizing the structure of this galaxy is the first S$\mathrm{\acute{e}}$rsic component with an index $n=1.27$, which is fairly small and less than the transition between classical bulges and pseudobulges at $n \approx 2$ \citep{Fisher:2008,Mathur:2011}.      
For the secondary S$\mathrm{\acute{e}}$rsic component, although the parameter determined ($n=0.10$) for the central excess might not be so robust, any larger index or a PSF model instead could hardly represent this sizable component.  Its physical scale (approximately 30~pc) is suggestive of a nuclear star cluster, as reported at the centers of a majority of spiral and lower-mass elliptical galaxies \citep{Boker:2002,Cote:2006}.   
While a large fraction of total magnitude was modeled with a low-brightness exponential disk, no spiral-like pattern was seen in a residual image.  Thus, Mrk~1388 comprises only spheroidal components.

The VLBA observation provides a very accurate astrometric position (Table~\ref{table:VLBA}) with an uncertainty $\lesssim1$~mas, which depends predominantly on the accuracy of the cataloged position of calibrator source in the International Celestial Reference Frame-2~(ICRF2; \citealt{Ma:2009}); 
$\Delta \mathrm{R.A.}=0.19$~mas and $\Delta \mathrm{Dec.}=0.34$~mas for J1455+2131.  The HST position of Mrk~1388 was apparently offset by \timeform{0.7''} north from that of VLBA.  However, the absolute astrometry information given in the FITS headers is typically inaccurate by approximately $\timeform{1''}$--$\timeform{2''}$ or worse.  Thus, the comparisons with radio-astrometric positions of VLA~(approximately $\timeform{0.1''}$) or VLBA~($\lesssim1$~mas) are meaningless.  No bright star or galaxy to refer an astrometric position against Mrk 1388 exists in the field of view of WFPC2.

\section{Discussion}\label{section:discussion}
The discussion herein focuses on the mass of the central black hole in Mrk~1388.  The black hole mass is one of key remaining parameters to discriminate between an unusual Sy2 and a Sy1 with an intermediate-mass black hole for Mrk~1388.   An example of VLBI detection exists in the Sy1 nucleus of NGC~4395 with an intermediate-mass black hole ($3.6 \times 10^5 \Mo$; \citealt{Filippenko:2003,Wrobel:2006}), which indicates nonthermal jets associated with a central engine of intermediate mass.     

The radio jet is relevant to mass accretion onto the central black hole.  A fundamental plane of black hole activity was discovered in three-dimensional (logarithmic) space comprising the nuclear radio luminosity $L_\mathrm{r}$ (in ergs~s$^{-1}$) at $5$~GHz, the X-ray luminosity $L_\mathrm{X}$ (in ergs~s$^{-1}$) at $2$--$10$~keV, and the black hole mass $M_\mathrm{BH}$ (in $\Mo$) ranging from $10^1$ to $10^{10}\Mo$ for samples of X-ray binaries, our Galactic Center, low-luminosity AGNs, Seyfert galaxies, and quasars: $\log{L_\mathrm{r}} = 0.60 \log{L_\mathrm{X}} + 0.78 \log{M_\mathrm{BH}} + 7.33$ \citep{Merloni:2003}.  This tendency of jet activity can be attributed to nonlinear dependencies on the black hole mass and the accretion rate \citep{Heinz:2003}.  Low-luminosity AGNs with low-accretion rates shows a relatively tighter correlation \citep{Kording:2006} and well-investigated by many studies \citep[e,g,][]{Yuan:2009,de-Gasperin:2011,Plotkin:2012}.  \citet{Gultekin:2014} showed the capability of the fundamental plane to estimate masses even for less than approximately $10^7\Mo$ at high accretion rates.  According to this relationship, the black hole mass of Mrk~1388 can be constrained by observed radio and X-ray emissions.

\begin{table}
\caption{Results of radial profile fit for HST image.\label{table:GALFIT}}
\begin{center}
\begin{footnotesize}
\begin{tabular}{lrrr}
\hline
\hline
Component&	Bulge &	Nuclear star cluster&	Exp.~disk \\
\hline
S$\mathrm{\acute{e}}$rsic index $n$&	1.27&	0.10&	\ldots \\
$R_\mathrm{e}$ or $R_\mathrm{s}$ (Pixel)\footnotemark[1]&	10.7&	1.45&	49.2 \\
\ \ \ \ \ \ \ \ \ \ \ \ \ \ \ \ \ (kpc)&	0.230&	0.031&	1.05 \\
P.A. (\timeform{D})&	$-22.8$&	$-16.4$&	$-27.0$ \\
Axis ratio ($b/a$)&	0.65&	0.87&	0.81 \\
Magnitude (mag)&	17.32&	17.73&	15.13 \\
\hline
\end{tabular}
\end{footnotesize}
\end{center}
\begin{flushleft}
{\footnotesize
\footnotemark[1] The effective radius $R_\mathrm{e}$ for the S$\mathrm{\acute{e}}$rsic profiles or the scale length of $R_\mathrm{s}$ ($=1.678R_\mathrm{e}$ for the $n=1$ case of a S$\mathrm{\acute{e}}$rsic profile) for the exponential disk profile.
}
\end{flushleft}
\end{table}

The ROSAT All-Sky Survey (ASS; $0.1$--$2.4$~keV) detected X-rays from Mrk~1388 with an intrinsic luminosity $L(0.5$--$2\ \mathrm{keV})=2.3\times10^{42}$~ergs~s$^{-1}$ and a photon index $\Gamma=1.8\pm2.2$ with an absorption column density $N_\mathrm{H}=9.6 \times 10^{20}$~cm$^{-2}$, assuming $H_0=50$~km~s$^{-1}$~Mpc$^{-1}$ \citep{Dennefeld:2003}.  We convert their result into the $H_0$-corrected 2--10~keV luminosity $L_\mathrm{X}=1.8\times10^{42}$~ergs~s$^{-1}$.  However, we must verify the poor statistics of the photon index $\Gamma$ (i.e., the absorption column density $N_\mathrm{H}$).  The ROSAT ASS Faint Source Catalog \citep{Voges:2000} provides the hardness ratios $HR1= 0.72\pm0.27$ (corresponding to $\Gamma=-1.4^{-0.5}_{-5.5}$ at $0.1$--$2.0$~keV) and $HR2=0.35\pm0.37$ ($\Gamma=-0.6^{+0.3}_{-1.8}$ at $0.5$--$2.0$~keV), which suggest a significant $N_\mathrm{H}$ with a negligible thermal component.  The negative detection of the broad component of Pa\emissiontype{$\beta$} \citep{Goodrich:1994,Veilleux:1997} indicates $A_\mathrm{V} \gtrsim11$~mag, which is equivalent to $N_\mathrm{H} \gtrsim 2 \times 10^{22}$~cm$^{-2}$ if we simply adopt the Galactic ratio $N_\mathrm{H}/A_\mathrm{V} \approx 1.8 \times 10^{21}$~cm$^{-2}$~mag$^{-1}$ \citep{Predehl:1995}.  Therefore, the ROSAT ASS luminosity presented above must be an underestimate, and should give a lower limit for the intrinsic X-ray luminosity.  On the other hand, the negative detection with the RXTE XSS of $<2.5\times10^{-11}$~ergs~cm$^{-2}$~s$^{-1}$ at a hard X-ray regime of 3--20~keV \citep{Heckman:2005} suggests $L(2$--$10\ \mathrm{keV})<2.3\times10^{43}$~ergs~s$^{-1}$.  % assuming $\Gamma=1.8$.  
By combining them, we obtain $1.8\times10^{42} < L(2$--10~$\mathrm{keV}) < 2.3\times10^{43}$~ergs~s$^{-1}$ as the possible range of X-ray luminosity.  

The absorption should be at most $N_\mathrm{H} = 10^{23}$~cm$^{-2}$, unless the ROSAT band ($0.1$--$2.4$~keV) is almost perfectly absorbed.  Thus, a plausible value is $2 \times 10^{22} < N_\mathrm{H} \lesssim 10^{23}$~cm$^{-2}$.  Furthermore, the ratio $L(2$--$10$~keV$)/L([\mathrm{O}\emissiontype{III}])$ \citep{Maiolino:1998} also indicates that $N_\mathrm{H}$ is not so large to be Compton-thick for Mrk~1388 as follows.  Using the intrinsic Balmer decrement $\mathrm{H\emissiontype{$\alpha$}/H\emissiontype{$\beta$}} = 3.1$, the observed values $13\times10^{-14}$ and $2.5 \times10^{-14}$~ergs~s$^{-1}$~cm$^{-2}$ for $\mathrm{H\emissiontype{$\alpha$}}$ and $\mathrm{H\emissiontype{$\beta$}}$, respectively \citep{Dahari:1988}, give $A_\mathrm{V}=1.5$~mag for the region of narrow-line components.  According to an observed [O\emissiontype{III}] flux \citep{Osterbrock:1985a} and an extinction correction at $\lambda5007$, we obtain $\log[L(2$--$10$~keV$)/L([\mathrm{O}\emissiontype{III}])]=1.4$, which is similar to typical values for Sy1 ($\approx1$--$2$; \citealt{Maiolino:1998}).  Thus, given range $2 \times 10^{22} < N_\mathrm{H} \lesssim 10^{23}$~cm$^{-2}$ of moderate absorption column density, the range of X-ray luminosity $L(2$--$10\ \mathrm{keV}) > 1.8\times10^{42}$~ergs~s$^{-1}$ inferred from the soft X-ray regime and the range $L(2$--$10\ \mathrm{keV}) < 2.3\times10^{43}$~ergs~s$^{-1}$ inferred from the hard X-ray regime are not so unreliable.   

According to the fundamental plane, the $2$--$10$~keV luminosity range and the 5-GHz luminosity $L_\mathrm{r}=8.6\times10^{37}$~ergs~s$^{-1}$ from our VLBI observation (Table~\ref{table:VLBA}) impose a black hole mass of $7.6\times10^5 < M_\mathrm{BH} < 5.4\times10^6 \Mo$.  This result includes the boundary between the intermediate-mass and the supermassive black hole near the lower end of the AGN mass function \citep{Zhou:2006,Greene:2007a,Doi:2012}.    

Alternatively, based on our HST image analysis (Section~\ref{section:HSTdata}), we also estimate the black hole mass using the relationship between the black hole mass and the S$\mathrm{\acute{e}}$rsic index $n$ (e.g., \citealt{Graham:2007}).  The S$\mathrm{\acute{e}}$rsic index of the major component in Mrk~1388 is $n=1.27$.  \citet{Graham:2007} presented $\log{M_\mathrm{BH}}= 7.98 + 3.70 \log{(n/3)} - 3.10 [\log{(n/3)}]^2$ with a total absolute scatter of $0.31$~dex from $27$~galaxies; this empirical relationship predicts a black hole mass $M_\mathrm{BH}=1.5\times10^6 \Mo$ for Mrk~1388, which is consistent with the value derived based on the fundamental plane discussed above.   

The third method we use is the improved version of the relationship between the black hole mass and the bulge luminosity \citep{Gultekin:2009}; namely, $\log{(M_\mathrm{BH}/\Mo)} = 8.95 + 1.11 \log{(L_\mathrm{V}/10^{11} L_{\solar V})}$ with an intrinsic scatter of $0.38 \pm 0.09$, where the $V$-band luminosity in solar-luminosity units is $\log{(L_\mathrm{V}/L_{\solar V})} = 0.4 (4.83 - M^0_{V\mathrm{bulge}})$ (see also \citealt{Graham:2013}).  The $V$-band absolute magnitude $M^0_{V\mathrm{bulge}}$ for the major S$\mathrm{\acute{e}}$rsic component with $n=1.27$ is $-17.4$~mag, which is derived based on the SDSS $g$--$r$ spectral index, $k$-correction, and Galactic extinction.  As a result, this $M_\mathrm{BH}$--$L$ relationship suggests a black hole mass $M_\mathrm{BH}=4.1\times10^6 \Mo$ for Mrk~1388.  This value is also consistent with those derived from the fundamental plane and the $M$--$n$ relationship discussed above.

Thus, all three methods provide similar estimates for the black hole mass; namely, $(0.76$--$5.4)\times10^6 \Mo$, $1.5\times10^6 \Mo$, and $4.1\times10^6 \Mo$.  These values are in a preferential range for typical NLS1 galaxies rather than for intermediate mass black holes.   
The detection of $FWHM(\mathrm{H\emissiontype{$\beta$}}) \sim 1200$--$1700$~km~s$^{-1}$ was expected from the empirical relationship between $R_\mathrm{BLR}$ and $L(2$--$10$~keV) \citep{Kaspi:2005}, where $R_\mathrm{BLR}$ is the radius of BLR, $M_\mathrm{BH}\approx v^2 R_\mathrm{BLR} / G$, and $v=(\sqrt{3}/2) FWHM$ \citep{Netzer:1990}, where $v$ is the effective velocity of BLR clouds and $G$ is the gravitational constant.  However, no broad-line component has yet been detected in H\emissiontype{$\alpha$}, H\emissiontype{$\beta$}, or Pa\emissiontype{$\beta$} (Section~\ref{section:introduction}).   
Thus, the hidden BLR is a promising idea.  
The picture of a heavily absorbed NLS1 nucleus can potentially explain the peculiarities previously observed (i.e., the negative detections of broad-line components in H\emissiontype{$\alpha$}, H\emissiontype{$\beta$}, or Pa\emissiontype{$\beta$} even with a respectable X-ray luminosity from an AGN and the strong high-ionization emission lines are observed), based on our radio observation and the analysis of galactic surface brightness, which are essentially not affected by dust extinction toward the nucleus.

The previous discovery of a hidden NLS1 galaxy in NGC~5506 is because of the detection of broad components (Pa\emissiontype{$\beta$}, O\emissiontype{I}$\lambda$1.1287~$\mu$m, etc.) in the near-infrared regime \citep{Nagar:2002}.  Because of a weaker extinction constraint ($A_\mathrm{V}>5$), the NGC~5506 nucleus may be less obscured than Mrk~1388.  NGC~5506 also shows other properties similar to Mrk~1388: [O\emissiontype{III}]$\lambda5007$/H\emissiontype{$\beta$}~$=7.5$, which is in the range of Sy2 galaxies, detection by VLBI with a high brightness temperature of $3.6\times10^8$~K \citep{Middelberg:2004}, and a moderate column density $N_\mathrm{H} = 3.4 \times 10^{22}$~cm$^{-2}$ \citep{Bassani:1999}.  
These sources suggest that obscured NLS1 galaxies might be hidden as a class in the range of the lowest-end of the AGN mass function or intermediate mass black holes.  A hard-X-ray imager with high sensitivity would discover numerous X-ray sources with rapid variations at the centers of pseudobulges.

\bigskip
% acknowledgment
We used the US National Aeronautics and Space Administration's (NASA) Astrophysics Data System~(ADS) abstract service and NASA/IPAC Extragalactic Database (NED), which is operated by the Jet Propulsion Laboratory~(JPL).  
The National Radio Astronomy Observatory is a facility of the National Science Foundation operated under cooperative agreement by Associated Universities, Inc. 
Based on an observation made with the NASA/ESA Hubble Space Telescope, obtained from the Data Archive at the Space Telescope Science Institute, which is operated by the Association of Universities for Research in Astronomy, Inc., under NASA contract NAS 5-26555.  These observations are associated with program \#5479.  
This study was partially supported by Grants-in-Aid for Scientific Research (B; 24340042, AD) from the Japan Society for the Promotion of Science (JSPS).

% \bibliography{mypaper}

\end{document}